\documentclass{aa}
\usepackage{graphicx}
\usepackage{supertabular}
\begin{document}
   \title{Southern Infrared Proper Motion Survey II: A sample of low
   mass stars with $\mu \geq 0.1''/ \rm{yr}$} 
   \titlerunning{Southern Infrared Proper Motion Survey II}
   \subtitle{}

   \author{N.R. Deacon
          \inst{1,2}
          \and N. C. Hambly\inst{2}   
          }
   \authorrunning{Deacon\& Hambly}
   \institute{\inst{1}Department of Astrophysics, Faculty of Science, Radboud University Nijmegen, P.O. Box 9010, 6500 GL Nijmegen, The Netherlands\\
  \inst{2}SUPA~\thanks{Scottish Universities' Physics Alliance}, Institute for Astronomy, University of
              Edinburgh, Royal Observatory Edinburgh, Blackford Hill, Edinburgh EH9 3HJ\\
              \email{ndeacon at astro.ru.nl,nd at roe.ac.uk}
             }

   \date{Received ---; accepted ---}

   \abstract{We present details of the second part of the Southern Infrared Proper Motion Survey (SIPS). Here accurate relative astrometry allows us to reduce the minimum proper motion to 0.1 arcseconds per year. This yields 6904 objects with proper motions between our minum cut and half an arcsecond a year. A small overspill sample with proper motions greater than this is also included. We examine our sample to identify interesting individual objects such as common proper motion binaries, potential L dwarfs and candidate nearby stars. Finally we show our survey is incomplete due to many factors, factors which we will take into account when simulating these survey results in the next paper in this series.   
   }

   \maketitle
%
\footnote{Full details of all objects identified by this survey are available in Tables 7 and 8 which are in an electronic format and can be obtained from CDS.}
\section{Introduction}
Proper motion surveys provide a wealth of information for the study of
Low Mass Stars and Brown Dwarfs. Firstly studies of the Luminosity
Function (and hence Mass Function, birthrate and space density) require large, clean samples of
cool objects. Proper motion surveys provide this by allowing the easy
exclusion of distant, intrinsically bright contaminants such
giants. Secondly they can also identify common proper motion
binaries. The make-up of these systems (and of their individual
components) can provide interesting insights, not only into
multiplicity, but into the star formation processes that created them
(Burgasser et al., 2005).

In Deacon, Hambly \& Cooke (2005) (hereafter DHC) we combined $J$, $H$, and $K_S$ data
from the 2MASS survey with $I_{N}$ data from SuperCOSMOS (Hambly et al., 2001) scans of UKST plates to
produce an infrared proper motion survey (SIPS) with a lower proper motion limit of half an arcsecond a year. Using such an
infrared proper motion survey allowed us to study low mass stars and
Brown Dwarfs in
the passbands in which they are brightest. This yielded
approximately 70 new high proper motion objects. While many of these
objects were interesting in themselves, it was clear that a reduction
in the lower proper motion limit (and hence an increase in the number
of objects detected) was required to produce a significant sample. Henced we have produced a survey
with a lower proper motion limit of $0.1$''/yr. For the sake of
comparison with SIPS I (DHC) we take the maximum proper motion of our
sample to be $0.5$/yr, although there will be a number of objects
which spill over this limit. 

The majority of current proper motion surveys have focussed on
identifying objects with high proper motions. Many, such as Lepine \& Shara
(2002), have proper motion limits well above our maximum proper motion
limit of $0.5$''/yr. Others (Scholz et al., 2002, Subasavage et al., 2005 and
Lepine, 2005) have lower limits of $0.45-0.4$''/yr, just encroching on
the region covered by our sample. Of those which go to lower limits
two, Ruiz et al. (2001) and Wroblewski \& Costa (2001), are limited in the
areas of sky they cover. This leaves only two surveys which cover the
majority of the southern hemisphere to lower proper motion limits
comparable to ours: Luyten's New Luten Two Tenths catalogue (NLTT, Luyten, 1979) and
Pokorny et al.'s 2003 Liverpool--Edinburgh High Proper Motion survey
(LEHPM). Both have lower proper motion limits of $0.2$''/yr,
however Luyten's survey is incomplete below $\delta=-30^\circ$. All of these surveys primarily make use of optical data.

Hence the approximately 7000 objects presented here provide a large
proper motion selected sample of low mass stars of comparable area and
depth to those currently available but with better completeness
for cool, red dwarfs. Here we present details of this sample along with a selection of interesting objects contained within it. In the third paper of this series we will outline the method used to simulate the survey results and the constraints on underlying distributions fundamental to star formation that can be set from these. 
\section{The SIPS Selection Method}
In Deacon, Hambly and Cooke (2005) we outlined the SIPS selection
method. Here we give a brief recap. 

The first stage of candidate selection is using 2MASS photometry. Here
objects are plotted on a $J-H$ vs. $H-K_s$ colour-colour diagram. An
object's position on this diagram leads it to be classified as a
potential M/L dwarf, early T dwarf or mid-late T dwarf. There is also
a fourth category in an overlap region between the mid M and mid T
dwarf range. Any object that does not fall into one of the four
categories is rejected. Next each candidates object had to be paired
with an $I$ plate counterpart. In the first run of the SIPS survey we
simply used the positions from both the UKST $I$ plates and the 2MASS
survey to calculate the movement of an object and hence its proper
motion. As the minimum proper motion in this case was half an
arcsecond per year this was generally well above a $5\sigma$
limit. However by reducing the lower proper motion limit to $0.1''/yr$
we run the risk of large errors in the measured proper motions and
therefore 
spurious detections. Hence we employ a relative error mapping
technique to reduce the RMS error on the proper motions. The details
of this are outlined in Section~\ref{RelAst}. Other than this the
candidate selection method is identical to that outlined in DHC with $I$ plate candidates being selected and
then filtered on $I-J$ colour, ellipticity, and being a good single
image far from bright stars. Following the initial selection process
all candidates were inspected by eye to reduce the number of spurious
detections. The proper motions were then calculated using the same
method as DHC. 
\subsection{Relative Astrometry}
\label{RelAst}
\begin{figure}[htb]     
        \begin{center}
\resizebox{\hsize}{!}{\includegraphics[scale=1.0]{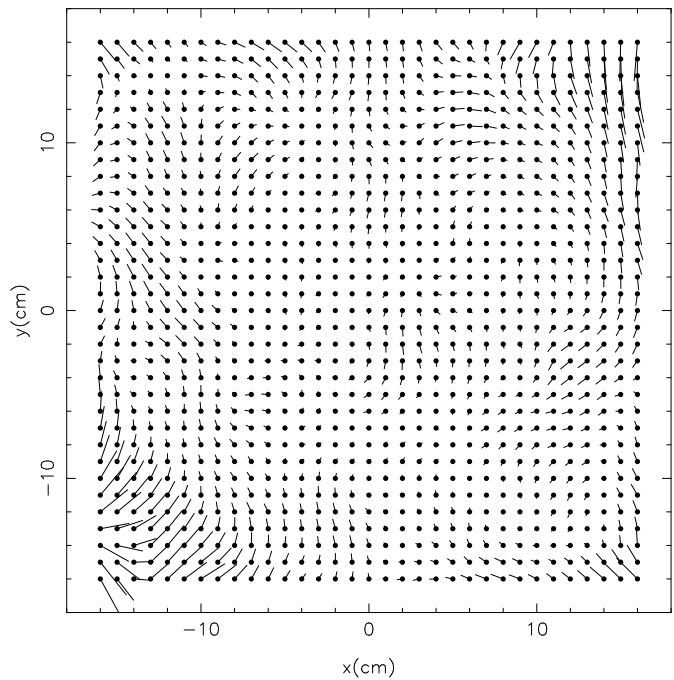}}
\end{center}
\caption{The errors across one single UKST I plate (field 411). The axes represent the position on the plate in centimetres and the size and direction of the lines on each lollipop show the positional offsets. The scale for the length of the lines is 4cm = 1''.}
         \label{swirl}
\end{figure}

In order to map out the systematic astrometric errors between 2MASS
and
the SSS, we employed the 2MASS catalogue positions as a standard.
Robust median offsets in X and Y in 1cm boxes were computed over the
field of view of each $I$ plate, and the residuals were smoothed and
filtered
within 3x3 boxes to create a map of any systematic offsets between
the photographic plate and 2MASS astrometry. A typical example of
the resulting systematic positional error map is shown in Figure~\ref{swirl},
with errors at the field edges of up to 0.5~arcsec. These positional
error
maps were then applied to the photographic positions before using
them, in conjunction with the 2MASS epoch positions, to measure
proper
motions.
\section{Results}
\begin{figure*}[htb]     
        \begin{center}
\resizebox{\hsize}{!}{\includegraphics[scale=0.9]{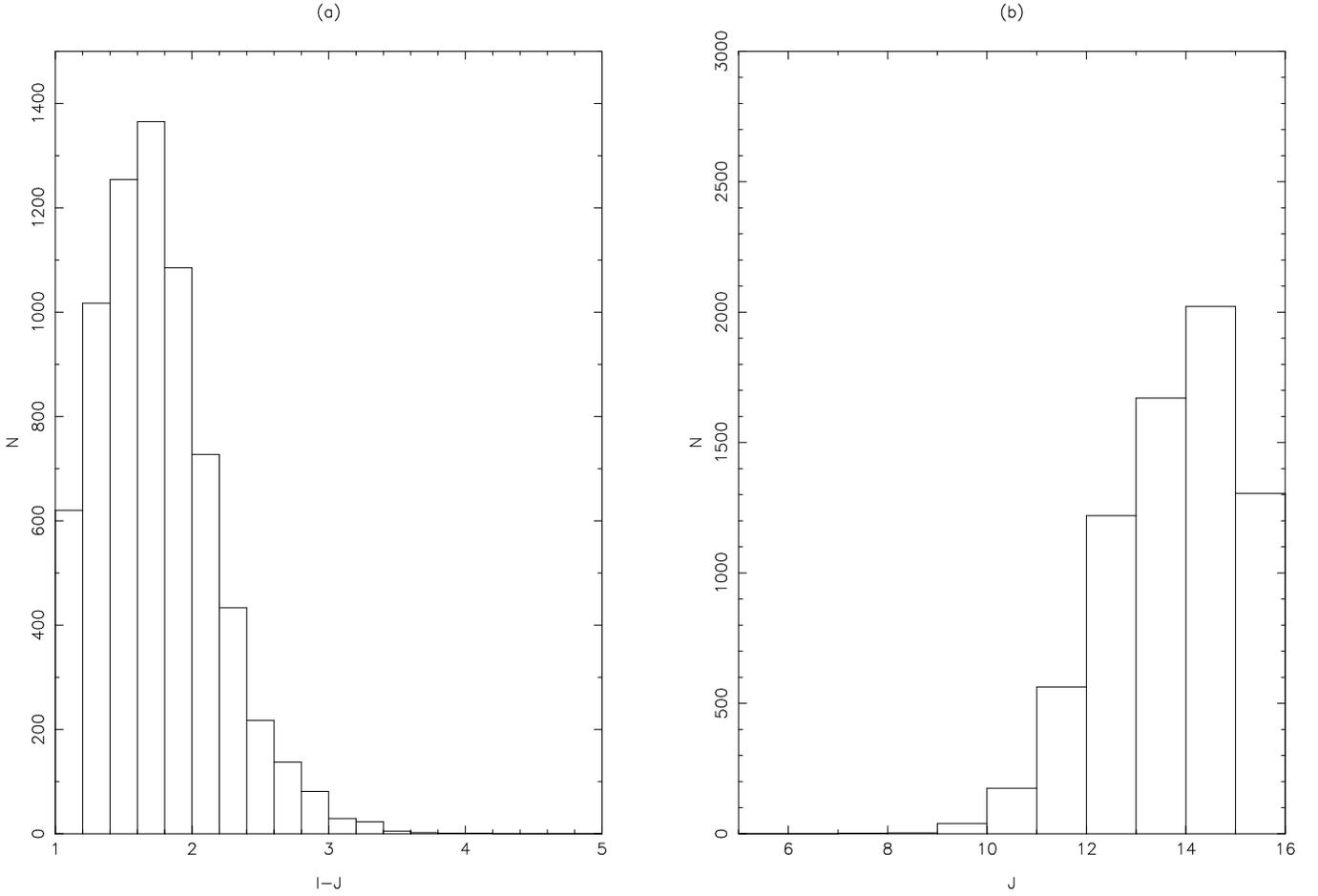}}
\end{center}
\caption{Histograms showing the magnitude ($J$) and colour ($I-J$) of
the objects found in this survey with proper motions between 0.1 and 0.5 arcseconds per year.}
         \label{IJhist}
\end{figure*}

The objects found in this survey with 0.1''/yr$< \mu <$0.5''/yr are shown in
Table 7. Figure~\ref{IJhist}(a) shows an $I-J$ histogram
for all objects in this sample, while Figure~\ref{IJhist}(b) shows
a histgram for $J$ magnitudes. It is clear that the objects make up a
large sample of mid-late M dwarfs.

\subsection{Higher Proper Motion Objects}
To prevent objects which had a proper motion below 0.5''/yr before the realtive astronometry correction, but above 0.5''/yr after it from slipping through the net, objects with proper motions just above our upper limit were also examined. Hence we also identified several objects with
proper motions greater than our upper limit of 0.5''/yr, these are
shown in Table 8. 

\subsection{Interesting Objects}
Rather than examine every object in detail we have sought to select
interesting individual objects, the details of which are listed in the
following three subsections.

\subsubsection{Red Objects}
In order to identify potential L and T dwarfs, we selected every
object found with $I-J>3.5$. The photometry and proper motions of
these objects are shown in Table~\ref{Redtab}. Comparing the $J-K_s$
colours with those for different spectral types in Kirkpatrick et
al. (2000) it becomes clear that SIPS2255-5713, SIPS0128-5545 and
SIPS0539-0059 are early-mid L dwarfs (ie. earlier than L6). In fact Fan et al. (2000)
spectroscopically identified that
SIPS0539-0059 (SDSS 0539-0059) is an L5 dwarf.
Additionally Reid et al. (in prep) find
that SIPS0719-5051 (2MASS J07193188-5051410) is an L0 dwarf. Kendall et al. (2006) identify SIPS2255-5713 and SIPS0128-5545 to be L5.5 and L1 respectively. It
appears from the $J, H, K_s$ colours that SIPS1625-2508 is a late M dwarf.
\begin{table*}
   \caption[]{Objects with $I-J>3.5$ found in this survey.Citation Key: $^1$ Kendall et al. (2006), $^2$ Reid et al. (in prep), $^3$ Fan et al. (2000).}
         \label{Redtab}
\begin{tabular}{lcccccc}
\hline
Name&\boldmath{$\mu$}&\boldmath{$I$}&\boldmath{$J$}&\boldmath{$H$}&\boldmath{$K_s$}&Previously\\
&(arcsecs/yr)&&&&&found in\\
\hline
SIPS2255-5713&0.341&17.85&14.08&13.19&12.58&$^1$\\
SIPS0128-5545&0.293&17.45&13.78&12.92&12.34&$^1$\\
SIPS0719-5051&0.141&17.60&14.09&13.28&12.77&$^2$\\
SIPS1625-2508&0.147&17.80&13.75&13.12&12.73&\\
SIPS0539-0059&0.363&17.84&14.03&13.10&12.53&$^3$\\
\hline
\end{tabular}
\end{table*}

We also compared object's $J$, $H$, and $K_S$ photometry with the
values which Kirkpatrick et al (2000) quote as typical for different
spectral types. All objects redder in $J-H$, $J-K_S$, $H-K_S$ than
Kirkpatrick's values for an M9 dwarf are listed in Table~\ref{Redtab1}. These objects are given spectral types based on their colours and those given in Kirkpatrick et al. (2000). SIPS2255-5713, SIPS0128-5545 and SIPS0539-0059 appear in thesamples in both Tables 1 and 2. 
\begin{table*}
   \caption[]{Objects with photometry suggesting they are L
   dwarfs. Our photometric spectral classifications are based on the
   colours in Kirkpatrick et al (2000). Citation key: $^1$ Kendall et al. (2006), $^2$ Cruz et al. (2003), $^3$ Delfosse et al (1999), $^4$ Kirkpatrick et al. (1999), $^5$ Dahn et al. (2002), $^6$ Fan et al. (2000).}
         \label{Redtab1}
\begin{tabular}{lcccccccc}
\hline
Name&\boldmath{$\mu$}&\boldmath{$I$}&\boldmath{$J$}&\boldmath{$H$}&\boldmath{$K_s$}&Photometric&Previous&Previously\\
&(arcsecs/yr)&&&&&Spectral Class&Spectral Class&found in\\
\hline
SIPS1753--6559&0.393&17.53&14.10&13.11&12.42&L2.5&&\\
SIPS2045--6332&0.223&15.95&12.62&11.81&11.21&L0&&\\
SIPS2255--5713&0.341&17.85&14.08&13.19&12.58&L1.5&L5.5&$^1$\\
SIPS0128--5545&0.293&17.45&13.78&12.92&12.34&L1&L1&$^1$\\
SIPS1341--3052&0.196&17.96&14.61&13.73&13.08&L2&&\\
SIPS2026--2943&0.417&17.78&14.80&13.95&13.36&L1&&\\
SIPS0316--2848&0.206&17.88&14.58&13.77&13.11&L1.5&L0&$^2$\\
SIPS0614--2019&0.308&17.90&14.78&13.90&13.38&L0.5&&\\
SIPS1058--1548&0.249&17.61&14.16&13.23&12.53&L2.5&L3$^4$&$^3$\\
SIPS1228--1547&0.415&17.62&14.38&13.35&12.77&L2&L5$^4$&$^3$\\
SIPS0847--1532&0.246&16.42&13.51&12.63&12.06&L1&L2&$^2$\\
SIPS0408--1450&0.256&17.44&14.22&13.34&12.82&L1&L2&$^2$\\
SIPS0058--0651&0.304&17.67&14.31&13.44&12.90&L1&L0$^2$&$^5$\\
SIPS0539--0059&0.363&17.84&14.03&13.10&12.53&L1.5&L5&$^6$\\
\hline
\end{tabular}
\end{table*}
\subsubsection{Common Proper Motion Objects}
\begin{figure}[htb]     
        \begin{center}
\resizebox{\hsize}{!}{\includegraphics[scale=0.9]{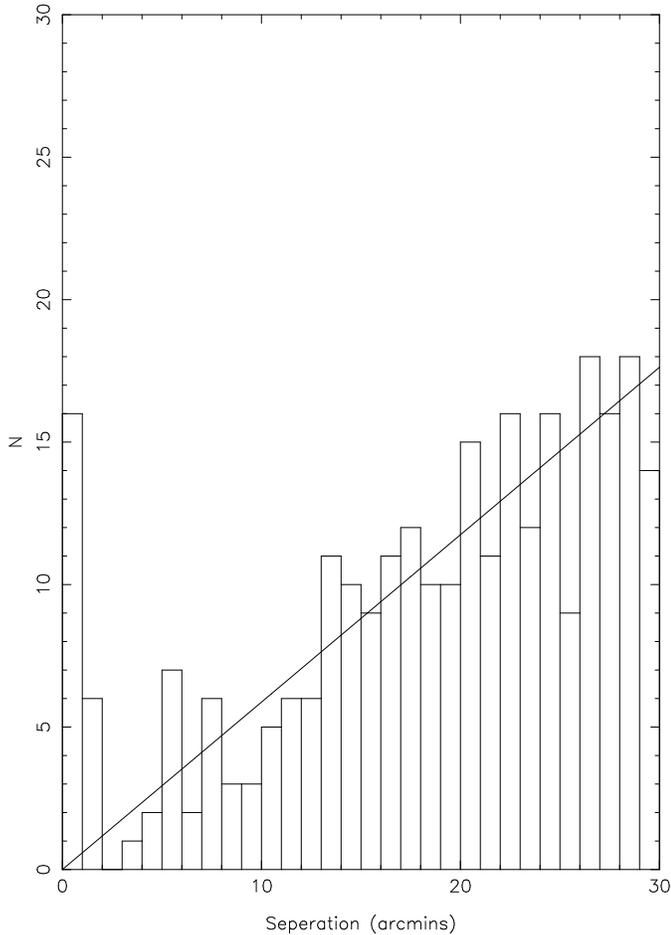}}
\end{center}
\caption{A histogram showing the seperations of objects with common
  proper motions in the SIPS survey. The line shows the expected
  distribution of randomly distributed objects, $dn \propto r dr$}
         \label{CPMhist}
\end{figure}

During the visual inspection phase of the data reduction process it
became apparent that there were many SIPS objects which shared a
common proper motion. However it is often difficult to distinguish
coincidence objects with the same proper motion from gravitiationally bound wide
binaries. To separate these two classes of objects we plotted the
separations of objects which had proper motions within $2\sigma$ of each
other. This is shown in Figure~\ref{CPMhist}. The straight line marks
the expected distribution of coincidence objects
randomly placed around the other object. Clearly the vast majority
follow this pattern. However the higher that expected number of pairs
with separations less than three arcminute indicates that a population
of binaries also contributes to this. Hence we choose a maximum
separation of three arcminutes for our binary sample. A list of SIPS
objects with separations less than this and proper motions within
$2\sigma$ of each other is shown in Table~\ref{CPMObs}.

Additionally during the cross-referencing process, several objects
which, while clearly not the target SIPS object, had a similar proper
motion to it were found. These were further investigated and any
companion found to have a SuperCOSMOS proper motion (Hambly et al., 2001b) differing by less than
$2\sigma$ from the SIPS object and to be closer than three arcminutes
to it is listed in Table~\ref{NLTTCPMObs}. Note that due to the manual
nature of this selection mechanism this list should not be regarded as
complete. 

In some cases it appears that the redder object of a particular pair is actually brighter than the bluer object. However closer examination reveals that these differences are not inconsistent with the typical photometric errors of $\sim 0.3$ magnitudes (Hambly et al., 2001c).
\begin{table*}
   \caption[]{SIPS objects which share common proper motions with
   other SIPS objects. PA = Position Angle. For NLTT objects see Luyten, for LEHPM objects see Pokorny et al. (2003), for WT objects see Wroblewski \& Torres (1991) and for $^1$ see Giclas, Burnham \& Thomas (1971). Pairs marked with * were identified as binaries by Luyten (1988). The pair marked with ** was found by Artigau et al. (2007).}
         \label{CPMObs}
\footnotesize
\begin{tabular}{lccrcccccl}
\hline
Name&Position&\boldmath{$\mu$}&PA&\boldmath{$\sigma_\mu$}&\boldmath{$I$}&\boldmath{$J$}&\boldmath{$H$}&\boldmath{$K_s$}&Other\\
&&(''/yr)&(\boldmath{$^\circ$})&(''/yr)&&&&&Name\\
\hline
\object{SIPS0650-7041}&06 50 59.75 -70 41 37.6&0.253&  2&0.025&16.85&14.67&14.00&13.51&\\
\object{SIPS0651-7041}&06 51 02.78 -70 41 38.8&0.253&  1&0.013&15.49&13.57&12.97&12.67&\\
\hline
\object{SIPS2241-5915b}&22 41 54.36 -59 15 30.8&0.323& 94&0.018&16.39&13.66&13.08&12.67&\\
\object{SIPS2241-5915a}&22 41 59.83 -59 15 12.4&0.334& 91&0.016&11.25&9.83&9.28&9.01& \object{LEHPM 5030}\\
\hline
\object{SIPS0630-5525a}&06 30 19.69 -55 25 48.2&0.254&137&0.007&13.71&12.42&11.87&11.59&\\
\object{SIPS0630-5525b}&06 30 23.39 -55 25 34.6&0.267&136&0.014&16.35&14.22&13.68&13.28&\\
\hline
\object{SIPS1917-5238b}&19 17 02.31 -52 38 47.0&0.254&181&0.016&11.26&9.94&9.35&9.08&\object{LDS 673b}*\\
\object{SIPS1917-5238a}&19 17 05.65 -52 38 49.8&0.258&182&0.014&11.40&10.19&9.64&9.31&\object{LDS 673a}*\\
\hline
\object{SIPS0358-5026}&03 58 15.41 -50 26 36.6&0.179& 34&0.022&16.73&14.69&14.10&13.75&\\
\object{SIPS0358-5027}&03 58 18.69 -50 27 39.8&0.179& 31&0.011&14.65&13.24&12.69&12.40&\\
\hline
\object{SIPS0126-5022}&01 26 55.26 -50 22 37.9&0.144&109&0.020&17.16&14.61&14.05&13.68&**\\
\object{SIPS0127-5023}&01 27 02.59 -50 23 20.2&0.170&112&0.023&17.62&14.81&14.16&13.62&**\\
\hline
\object{SIPS1131-4419}&11 31 00.17 -44 19 08.9&0.111&269&0.011&12.11&11.03&10.40&10.15&\\
\object{SIPS1131-4418}&11 31 01.35 -44 18 54.4&0.120&267&0.009&14.04&12.69&12.05&11.77&\\
\hline
\object{SIPS0207-4052b}&02 07 54.69 -40 52 16.4&0.255&206&0.016&12.45&11.36&10.82&10.55&\object{NLTT7110}*\\
\object{SIPS0207-4052a}&02 07 55.77 -40 52 30.9&0.251&209&0.015&12.53&11.49&10.96&10.66&\object{NLTT7107}*\\
\hline
\object{SIPS0116-4005b}&01 16 13.97 -40 05 02.1&0.170& 96&0.011&14.06&12.56&11.98&11.69&\\
\object{SIPS0116-4005a}&01 16 15.44 -40 05 12.1&0.157& 95&0.011&16.10&14.36&13.78&13.40&\\
\hline
\object{SIPS2145-3612}&21 45 49.39 -36 12 25.9&0.121&151&0.020&13.47&12.42&11.81&11.56&\\
\object{SIPS2145-3610}&21 45 54.45 -36 10 55.8&0.119&153&0.021&15.56&14.26&13.66&13.41&\\
\hline
\object{SIPS0252-3438a}&02 52 32.11 -34 38 49.0&0.308& 86&0.011&14.04&12.53&11.93&11.63&\object{LEHPM 2856}*\\
\object{SIPS0252-3438b}&02 52 33.26 -34 38 54.3&0.303& 86&0.011&13.69&12.25&11.68&11.38&\object{LEHPM 2857}*\\
\hline
\object{SIPS0118-2730}&01 18 10.69 -27 30 39.0&0.117& 90&0.011&15.86&13.90&13.27&12.95&\\
\object{SIPS0118-2729}&01 18 06.58 -27 29 03.3&0.115& 85&0.014&13.94&12.45&11.90&11.61&\\
\hline
\object{SIPS0215-2440}&02 15 13.69 -24 40 06.2&0.177& 89&0.010&14.89&13.23&12.61&12.36&\object{LDS 3363b}*\\
\object{SIPS0215-2439}&02 15 15.02 -24 39 43.6&0.181& 90&0.010&14.14&12.54&11.97&11.66&\object{LDS 3363a}*\\
\hline
\object{SIPS1018--2028a}&10 18 12.27 -20 28 22.1&0.397&285&0.010&12.21&10.59&10.01&9.71&NLTT23954*\\
\object{SIPS1018--2028b}&10 18 14.01 -20 28 41.9&0.407&286&0.013&10.28&9.00&8.42&8.15&NLTT23956*\\
\hline
\object{SIPS0427-1548}&04 27 20.96 -15 48 33.2&0.170&111&0.021&17.19&14.48&13.71&13.28&\\
\object{SIPS0427-1547}&04 27 22.20 -15 47 59.0&0.160&112&0.009&13.48&12.07&11.48&11.17&\\
\hline
\object{SIPS0229-1540}&02 29 44.57 -15 40 34.6&0.156&122&0.030&15.86&13.86&13.29&12.98&\\
\object{SIPS0229-1541}&02 29 46.98 -15 41 47.6&0.145&127&0.011&16.11&14.32&13.72&13.35&\\
\hline
\object{SIPS0441-1356}&04 41 58.63 -13 56 05.4&0.292& 86&0.028&12.63&11.43&10.87&10.60&\object{NLTT 13776}\\
\object{SIPS0442-1356}&04 42 00.44 -13 56 23.7&0.278& 77&0.034&15.21&12.98&12.36&11.97&\\
\hline
\object{SIPS0116-1318a}&01 16 48.12 -13 18 19.4&0.130&103&0.021&15.19&14.16&13.57&13.32&\\
\object{SIPS0116-1318b}&01 16 49.11 -13 18 55.0&0.133&105&0.021&16.42&14.93&14.41&14.07&\\
\hline
\object{SIPS1402-0312}&14 02 22.81 -03 12 16.9&0.401&176&0.020&12.55&11.52&10.99&10.68&\object{NLTT 36053}*\\
\object{SIPS1402-0311}&14 02 24.01 -03 11 55.5&0.393&175&0.021&12.20&11.08&10.57&10.28&\object{G 64-36}$^1$*\\
\hline
\object{SIPS0820-0231}&08 20 12.08 -02 31 09.2&0.202&166&0.031&13.24&12.13&11.60&11.29&\object{LDS 3786a}*\\
\object{SIPS0820-0230}&08 20 12.74 -02 30 59.5&0.170&174&0.034&13.65&12.33&11.79&11.45&\object{LDS 3786b}*\\
\hline
\object{SIPS0005-0139}&00 05 36.22 -01 39 39.7&0.331& 67&0.015&14.30&12.88&12.35&12.06&\object{NLTT 175}*\\
\object{SIPS0005-0139}&00 05 36.73 -01 39 57.2&0.336& 66&0.015&13.00&11.86&11.31&11.07&\object{NLTT 176}*\\
\hline
\object{SIPS2315-0045a}&23 15 43.90 -00 45 00.8&0.124& 84&0.009&14.34&12.46&11.88&11.58&\object{LDS 6019a}*\\
\object{SIPS2315-0044b}&23 15 46.52 -00 44 06.5&0.130& 84&0.009&14.69&12.81&12.20&11.92&\object{LDS 6019b}*\\
\hline
\end{tabular}
\normalsize
\end{table*}
\begin{table*}
   \caption[]{SIPS objects which share common proper motions with
   objects found in other studies which are not themselves SIPS objects. These pairings were found during the cross referencing process and should not be regarded as a complete list. PA = Position Angle. For LEHPM objects see
   Pokorny (2003), for NLTT objects see NLTT, $^1$ see
   Lasker et al. (1990), $^2$ see Gizis et al. (2000) and $^3$ see Schonfeld (1886). All pairs marked * were identified as common proper motion pairs by Luyten (1988)}
         \label{NLTTCPMObs}
\footnotesize
\begin{tabular}{lccrcl}
\hline
Name&Position&$\mu$&PA&$\sigma_\mu$&Other\\
&&(''/yr)&($^\circ$)&(''/yr)&Name\\
\hline
\object{SIPS0551-8116}&05 51 54.60 -81 16 09.6&0.233&20&0.009&\\
\object{NLTT 15903}&05 52 38.16 -81 16 03.0&0.223&18&0.016&\\
\hline
\object{SIPS2126-7337}&21 26 20.62 -73 37 10.1&0.176&83&0.010&\\
\object{GSC 09334-00112}$^1$&21 26 30.94 -73 38 23.2&0.184&85&0.016&\\
\hline
\object{SIPS1954-6117}&19 54 51.61 -61 17 05.5&0.214&163&0.033&\\
\object{NLTT 48361}&19 54 49.14 -61 19 19.1&0.190&177&-&\\
\hline
\object{SIPS0447-5823}&04 47 11.69 -58 23 21.0&0.279&30.263&0.029&\\
\object{LEHPM 3838}& 04 47 13.38 -58 23 20.6&0.265&31.4&0.080&WT 155\\
\hline
\object{SIPS1858-4513}&18 58 21.39 -45 13 33.8&0.231&124&0.014&\\
\object{NLTT 47218}&18 58 16.68 -45 14 12.6&0.254&120&0.028&\\
\hline
\object{SIPS0333-4324}&03 33 18.57 -43 24 56.9&0.316&59&0.025&\\
\object{NLTT 11245}&03 33 18.03 -43 25 12.0&0.281&58&0.018&\\
\hline
\object{SIPS0123-3507}&01 23 24.56 -35 07 18.4&0.213&220&0.013&\object{LEHPM 1491}\\
\object{NLTT 4642}&01 23 29.93 -35 08 27.7&0.223&220&0.014&\object{LEHPM 1492}\\
\hline
\object{SIPS0116-3342}&01 16 31.19 -33 42 51.3&0.189&42&0.017&\object{LDS 3257b}*\\
\object{NLTT 4263}&01 16 32 78 -33 42 54.8&0.185&44&0.014&*\\
\hline
\object{SIPS0933-2752}&09 33 37.63 -27 52 47.4&0.328&297&0.022&\\
\object{NLTT 22073}&09 33 36.22 -27 52 25.0&0.311&289&0.017&\\
\hline
\object{SIPS1917-2748}&19 17 18.10 -27 48 54.0&0.209&133&0.028&\object{LDS 4807b}*\\
\object{NLTT 47615}&19 17 22.63 -27 47 32.26&0.195&141&0.008&*\\
\hline
\object{SIPS0132-2744}&01 32 35.91 -27 44 37.1&0.195&128&0.013&\object{NLTT 5135}*\\
\object{NLTT 5136}&01 32 37.48 -27 44 25.4&0.191&129&0.016&*\\
\hline
\object{SIPS0028-2651}&00 28 10.24 -26 51 24.7&0.189&208&0.015&\object{NLTT 1492}*\\
\object{NLTT 1491}&00 28 08.06 -26 52 25.8&0.198&207&0.021&*\\
\hline
\object{SIPS2147-2644}&21 47 44.64 -26 44 05.4&0.235&215&0.022&\object{2MASSW J2147446-264406}$^2$\\
\object{NLTT 52094}&21 47 47.00 -26 42 52.7&0.252&211&0.012&\\
\hline
\object{SIPS0227-2630}&02 27 25.50 -26 30 07.8&0.165&56&0.012&\object{NLTT 8057}*\\
\object{NLTT 8059}&02 27 24.37 -26 29 42.1&0.159&60&0.022&*\\
\hline
\object{SIPS1135-2017}&11 35 05.43 -20 17 26.3&0.245&279&0.017&\object{NLTT 27883}*\\
\object{NLTT 27884}&11 35 05.03 -20 16 57.6&0.231&281&0.024&*\\
\hline
\object{SIPS0133-1948}&01 33 08.77 -19 48 34.3&0.307&65&0.021&\object{NLTT 5169}*\\
\object{NLTT 5163}&01 33 05.55 -19 50 22.7&0.322&67&0.022&*\\
\hline
\object{SIPS1301-1848}&13 01 51.64 -18 48 40.5&0.335&286&0.016&\object{NLTT 32629}*\\
\object{NLTT 32636}&13 01 42.64 -18 47 23.2&0.321&287&0.021&*\\
\hline
\object{SIPS0532-1605}&05 32 33.82 -16 05 53.2&0.272&186&0.023&\\
\object{NLTT 15268}&05 32 42.06 -16 06 00.3&0.289&189&0.025&\\
\hline
\object{SIPS0536-1302}&05 36 07.79 -13 02 09.4&0.199&162&0.013&\object{NLTT 15357}*\\
\object{NLTT 15358}&05 36 08.58 -13 02 40.1&0.202&174&0.025&*\\
\hline
\object{SIPS2016-1100}&20 16 49.26 -11 00 13.7&0.349&206&0.019&\object{NLTT 48967}*\\
\object{NLTT 48973}&20 16 55.72 -10 58 54.6&0.318&206&-&*\\
\hline
\object{SIPS0321-0807}&03 21 46.26 -08 07 13.7&0.118&72&0.014&\\
\object{BD-08 638}$^3$&03 21 48.90 -08 06 10.58&0.127&76&0.019&\\
\hline
\object{SIPS1402-0447}&14 02 14.40 -04 47 53.2&0.231&254&0.018&\object{NLTT 36042}*\\
\object{NLTT 36043}&14 02 14.90 -04 48 09.0&0.214&250&0.021&*\\
\hline
\object{SIPS0023-0342}&00 23 30.92 -03 42 20.4&0.189&52.79&0.028&\object{NLTT 1210}*\\
\object{NLTT 1213/1214}&00 23 32.29 -03 42 28.4&0.232&55&-&*\\
\hline
\end{tabular}
\normalsize
\end{table*}
\subsubsection{Potential nearby stars}
In order to identify nearby objects within the sample, we estimated the distances of objects using the RECONS colour-absolute magnitude relations (Henry et al., 2004). These relations allowed us to calculate absolute $K_S$ magnitudes of a star from the $I-J$, $I-H$ and $I-K_S$ colours. Each of these estimates could then be combined with the apparent magnitude to yield a distance modulus (and hence distance) for each star. However we needed to gain a clear picture of the potential errors in such a calculation. Hence we  used these relations to calculate the distance modulii for the sample of nearby stars compiled by Reid using photometry from Leggett (1992) and Bessell (1991) (with an additional simulated error on the $I$ band data to mimic the less accurate plate photometry) and compared them with those distance modulii calculated from their trigonometric parallaxes. We found that these three distance relations when combined produced distance modulii that were 0.3 magitudes too close, with each having an error of one magnitude. This error is then combined with the random error from each measurement to produce an estimate for the total error on the distance (after the 0.3 magnitude offset was removed) of each star.

The calculated distance estimates are shown in Table~\ref{close0.2} for the sample with 0.1''/yr$< \mu <$0.5''/yr and in Table~\ref{close0.5} for those objects with proper motions above 0.5''/yr. In total there are 12 stars with distance estimates closer than 20pc which have not been found before.
\footnotesize
\begin{table*}   
\caption[]{Objects in this sample with proper motions in the range 0.1''/yr<$\mu$<0.5''/yr which have estimated distances closer that 20pc. Citation key: $^1$ Subasavage et al. (2005), $^2$ Kendall et al. (2006), $^3$ Webb et al. (1999), $^4$ Reyle et al. (2002), $^5$ Phan Bao et al. (2003), $^6$ Giclas, Burnham \& Thomas (1971), $^7$ Gizis et al. (2002), $^8$ Fan et al. (2000). For objects designated NLTT or LP see Luyten's New Luyten Two Tenths catalogue (1979), for objects marked CE see Ruiz et al. (2001).}
         \label{close0.2}
\begin{supertabular}{lcccccccl}
\hline
Name&Position&Distance&$\mu$&$I$&$J$&$H$&$K_S$&Other\\
&&pc&(''/yr)&&&&&Name\\
\hline
\object{SIPS1848-8214} &18 48 51.35 -82 14 40.4&14.26$\pm^{8.50}_{5.33}$&0.272&14.373&11.482&10.922&10.503&\\ 
\object{SIPS1731-7851} &17 31 45.16 -78 51 23.3&16.94$\pm^{10.09}_{6.32}$&0.362&14.173&11.613&11.007&10.675&\\ 
\object{SIPS0630-7643} &06 30 46.68 -76 43 15.3&8.69$\pm^{5.18}_{3.25}$&0.483&10.738&8.894&8.275&7.923&\object{SCR J0630-7643} $^1$\\ 
\object{SIPS1809-7613} &18 09 06.96 -76 13 23.0&15.92$\pm^{9.55}_{5.97}$&0.156&11.615&9.817&9.275&8.989&\\ 
\object{SIPS2016-7531} &20 16 10.88 -75 31 04.8&19.06$\pm^{11.38}_{7.12}$&0.253&12.247&10.465&9.864&9.509&\\ 
\object{SIPS0504-7401} &05 04 26.15 -74 01 55.3&19.03$\pm^{11.36}_{7.12}$&0.359&12.147&10.346&9.778&9.469&\\ 
\object{SIPS1826-6542} &18 26 46.79 -65 42 37.7&11.28$\pm^{6.72}_{4.21}$&0.311&12.913&10.569&9.960&9.547&\\ 
\object{SIPS2045-6332} &20 45 02.28 -63 32 05.3&15.15$\pm^{9.03}_{5.66}$&0.223&15.950&12.619&11.807&11.207&\\ 
\object{SIPS0152-6329} &01 52 55.17 -63 29 30.2&16.67$\pm^{9.94}_{6.23}$&0.140&11.993&10.167&9.604&9.261&\\ 
\object{SIPS2019-5816} &20 19 49.82 -58 16 41.1&13.66$\pm^{8.14}_{5.10}$&0.347&12.868&10.664&10.104&9.715&\\ 
\object{SIPS2255-5713} &22 55 18.70 -57 13 04.0&12.92$\pm^{8.72}_{5.21}$&0.341&17.851&14.083&13.189&12.579&\object{2M2255-57} $^2$\\ 
\object{SIPS0128-5545} &01 28 26.76 -55 45 34.6&15.24$\pm^{9.79}_{5.96}$&0.293&17.447&13.775&12.916&12.336&\object{2M0128-55} $^2$\\ 
\object{SIPS0139-3936} &01 39 21.55 -39 36 05.8&14.99$\pm^{8.96}_{5.61}$&0.283&10.791&9.209&8.629&8.274&\\ 
\object{SIPS1110-3731} &11 10 27.98 -37 31 51.8&10.72$\pm^{6.41}_{4.01}$&0.103&9.041&7.651&7.041&6.774&\object{TWA 3A} $^3$\\ 
\object{SIPS0339-3525} &03 39 34.82 -35 25 48.4&9.95$\pm^{5.96}_{3.73}$&0.405&13.288&10.725&10.017&9.548&\object{APMPM J0340-3526}  $^4$\\ 
\object{SIPS0604-3433} &06 04 52.13 -34 33 38.6&13.76$\pm^{8.26}_{5.16}$&0.330&9.027&7.742&7.183&6.866&\object{LP 949-15}\\ 
\object{SIPS0124-3355} &01 24 30.52 -33 55 00.6&18.85$\pm^{11.28}_{7.06}$&0.222&12.445&10.555&10.009&9.682&\object{LP 939-44}\\ 
\object{SIPS1346-3149} &13 46 46.24 -31 49 26.7&16.03$\pm^{9.58}_{6.00}$&0.366&13.165&10.975&10.436&10.038&\object{LP 911-56}\\ 
\object{SIPS1037-2746} &10 37 45.41 -27 46 40.2&18.13$\pm^{10.86}_{6.79}$&0.322&9.936&8.595&8.035&7.719&\object{CE 186}\\ 
\object{SIPS2351-2537} &23 51 50.25 -25 37 37.8&18.35$\pm^{10.94}_{6.86}$&0.420&15.573&12.471&11.725&11.269&\object{LEHPM 6334}\\ 
\object{SIPS1625-2508} &16 25 49.66 -25 08 17.8&8.62$\pm^{16.80}_{5.70}$&0.147&17.800&13.745&13.117&12.728&\\ 
\object{SIPS1042-2416} &10 42 41.32 -24 16 06.2&17.13$\pm^{10.21}_{6.40}$&0.205&12.087&10.277&9.672&9.338&\object{NLTT 25128}\\ 
\object{SIPS1504-2355} &15 04 16.35 -23 55 55.8&19.12$\pm^{11.41}_{7.15}$&0.327&14.652&12.011&11.383&11.032&\object{LP 859-1}\\ 
\object{SIPS1309-2330} &13 09 21.85 -23 30 33.9&15.64$\pm^{9.35}_{5.85}$&0.383&14.514&11.785&11.082&10.669&\object{CE 303}\\ 
\object{SIPS1155-2224} &11 55 42.94 -22 24 58.2&13.50$\pm^{8.05}_{5.04}$&0.369&13.203&10.930&10.295&9.881&\object{DENIS J115542.9-222458} $^5$\\ 
\object{SIPS1351-1758} &13 51 57.20 -17 58 49.1&18.18$\pm^{10.85}_{6.80}$&0.222&12.815&10.823&10.210&9.927&\object{LP 798-49}\\ 
\object{SIPS0931-1717} &09 31 22.41 -17 17 41.8&17.46$\pm^{10.43}_{6.53}$&0.359&13.140&11.073&10.467&10.069&\object{DENIS J093122.3-171742} $^5$\\ 
\object{SIPS0435-1607} &04 35 15.97 -16 07 02.0&14.25$\pm^{8.55}_{5.34}$&0.355&12.341&10.406&9.779&9.352&\object{LP 775-31}\\ 
\object{SIPS0440-0530} &04 40 23.32 -05 30 07.8&10.21$\pm^{6.10}_{3.82}$&0.356&13.171&10.658&9.986&9.545&\object{LP 655-48}\\ 
\object{SIPS1324-0504} &13 24 46.44 -05 04 17.7&17.18$\pm^{10.25}_{6.42}$&0.328&11.043&9.465&8.861&8.563&\object{G 14-52} $^6$\\ 
\object{SIPS1607-0442} &16 07 31.25 -04 42 06.3&15.06$\pm^{8.99}_{5.63}$&0.462&14.866&11.896&11.187&10.717&\object{2MASSW J1607312-044209} $^7$\\ 
\object{SIPS0109-0343} &01 09 51.04 -03 43 26.3&12.74$\pm^{7.60}_{4.76}$&0.376&14.768&11.694&10.931&10.428&\object{LP 647-13}\\ 
\object{SIPS1712-0323} &17 12 04.49 -03 23 28.9&15.13$\pm^{9.02}_{5.65}$&0.385&14.482&11.607&10.994&10.637&\\ 
\object{SIPS1614-0251} &16 14 25.20 -02 51 03.5&16.73$\pm^{9.98}_{6.25}$&0.350&13.528&11.303&10.683&10.280&\object{LP 624-54}\\ 
\object{SIPS0539-0059} &05 39 51.86 -00 59 05.2&11.29$\pm^{7.68}_{4.57}$&0.363&17.836&14.033&13.104&12.527&\object{SDSS J053951.9-005901} $^8$\\ 
\hline
\end{supertabular}
\end{table*}
\normalsize
\begin{table*}
   \caption[]{Objects in this sample with proper motions above 0.5''/yr which have estimated distances closer that 20pc. Citation key: $^1$ Gizis et al. (2000), $^2$ Luyten Half arcsecond Catalogue (1979), $^3$ Gliese \& Jahreiss (1979).}
         \label{close0.5}
\footnotesize
\begin{tabular}{lcccccccl}
\hline
Name&Position&Distance&$\mu$&$I$&$J$&$H$&$K_S$&Other\\
&&pc&(''/yr)&&&&&Name\\
\hline
SIPS1241-3843 &12 41 08.28 -38 43 11.0&18.53$\pm^{11.06}_{6.93}$&0.506&13.654&11.477&10.825&10.450&\object{2MASSW J1241080-384312} $^1$\\ 
SIPS1552-2623 &15 52 44.51 -26 23 10.7&12.98$\pm^{7.73}_{4.84}$&0.525&12.313&10.258&9.676&9.315&\object{LHS 5303} $^2$\\ 
SIPS0853-0329 &08 53 36.38 -03 29 30.8&11.30$\pm^{6.77}_{4.23}$&0.708&13.920&11.212&10.469&9.942&\object{GJ 3517 }$^3$\\
\hline
\end{tabular}
\normalsize
\end{table*}
\subsection{{\bf Completeness}}
The final aim of this survey is to study the
local space density, mass function and birthrate of cool dwarfs. If we are to properly examine these we must first estimate the completeness of the survey. Firstly there is the problem of crowded regions. These were excluded by utilising the proximity flag in the 2MASS data files. There will also be problems relating to the UKST images. The SuperCOSMOS software flags objects which are blended with other objects. These have been excluded along with those falling near bright stars. The incompletenesses caused by these three effects have been examined and are quantified in Section 5 of DHC.

\begin{figure}[htb]     
        \begin{center}
\resizebox{\hsize}{!}{\includegraphics[scale=0.9]{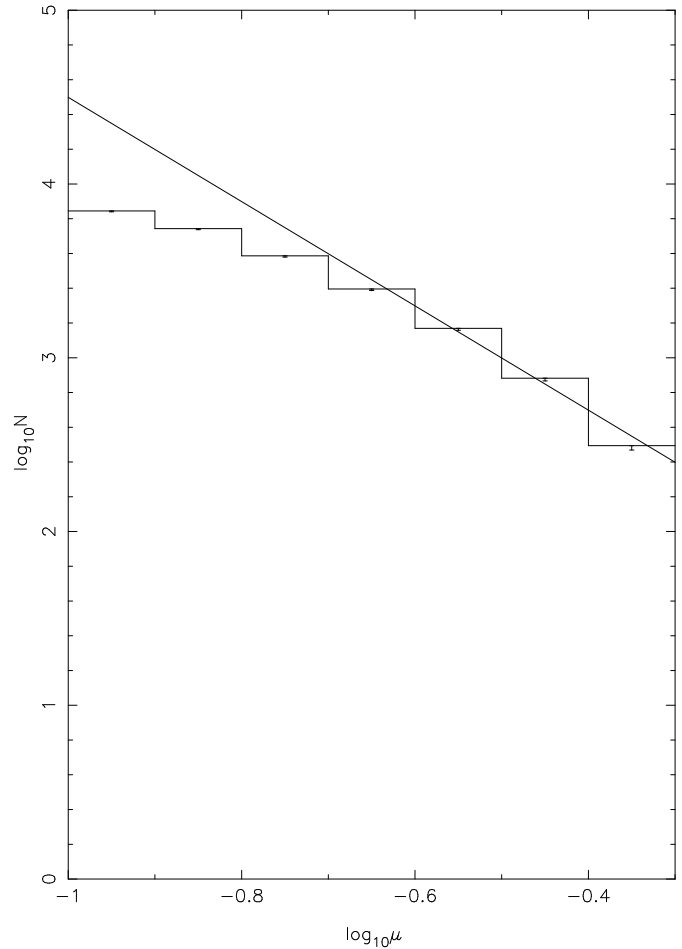}}
\end{center}
\caption{The cumulitive number of objects with proper motions greater
  than the minimum limit of each bin. The solid line shows the $N
  \propto \mu^-3$ trend which would be expected for a totally complete
  sample. It is clear that the sample is incomplete below 0.2''/yr. The proper motion ($\mu$) is in arcseconds/year.}
         \label{PMhist}
\end{figure}

There will also be incompleteness caused by both the limiting
magnitudes of the survey and the short epoch separation on some
plates. To illustrate this a histogram of the cumulative number of
objects with proper motions greater than the minimum proper motion in
each bin is plotted in Figure~\ref{PMhist}. If the survey was totally
complete it would be expected that the number of objects in each bin
would scale as $\mu^{-3}$ (see the solid line). However it is clear
that this is not the case and that the incompleteness is significant below $0.2$''/yr.

In order to gain information on the mass function and birthrate of
cool dwarfs all the sources of incompleteness mentioned in this
section must be taken into account. In the third paper of this series we will detail the
simulations which use both the crowding incompleteness calculated in DHC
and the other selection effect to produce simulated samples. These will
then used to constrain underlying distributions such as the birthrate and mass function. 
\section{Conclusions}
Here we have presented a large sample of cool low mass objects. Within this sample we have identified 45 common proper motion systems of which 11 had neither candidate previously identified. In addition 38 objects (12 of them new) which may lie within 20pc have been found, along with 4 new potential L dwarfs. It has been shown that the surevy is incomplete and that the sources of incompleteness must be taken into account before examining the constraints that can be set on the mass function and stellar birthrate.

In the next paper in this series we will model the low mass stellar population. This will allow us to take into account all the selection criteria and their effects, along with the various sources of incompleteness. This will allow us to constrain the underlying mass function and birthrate. 
\begin{acknowledgements}The authors would like to thank Sue Tritton and Mike Read for their
help in selecting plates, Harvey MacGillivray and Eve Thomson for
their scanning of the material on SuperCOSMOS and Todd Henry and John Cooke for their helpful discussions. 
This publication makes use of data products from the Two Micron All Sky Survey, which is a joint project of the University of
Massachusetts and the Infrared Processing and Analysis Center/California Institute of Technology, funded by the National
Aeronautics and Space Administration and the National Science Foundation. SuperCOSMOS wass funded by a grant from the
UK Particle Physics and Astronomy Research Council. This publication makes use of the SLALIB positional astronomy library (Wallace, 1998).     
\end{acknowledgements}

\end{document}